\documentclass[doublecol]{epl2}

\title{Bounds of efficiency at maximum power for linear, superlinear and sublinear irreversible Carnot-like heat engines}
\shorttitle{Bounds of EMP for three irreversible Carnot-like Engines}

\author{ Yang Wang \and Z. C. Tu\thanks{Corresponding author. E-mail: \email{tuzc@bnu.edu.cn}}}

\shortauthor{Wang and Tu}

\institute{Department of Physics, Beijing Normal University, Beijing 100875, China}
\pacs{05.70.Ln}{Nonequilibrium and irreversible thermodynamics}

\abstract{The efficiency at maximum power (EMP) of irreversible Carnot-like heat engines is investigated based on the weak endoreversible assumption and the phenomenologically irreversible
thermodynamics. It is found that the weak endoreversible assumption can reduce to the conventional one for the heat engines working at maximum power. Carnot-like heat engines are classified into three types (linear, superlinear, and sublinear) according to different characteristics of constitutive relations between the heat transfer rate and the thermodynamic force. The EMPs of Carnot-like heat engines are proved to be bounded between $\eta_C/2$ and $\eta_C/(2-\eta_C)$ for the linear type, 0 and $\eta_C/(2-\eta_C)$ for the superlinear type, and $\eta_C/2$ and $\eta_C$ for the sublinear type, respectively, where $\eta_C$ is the Carnot efficiency.
}

\begin{document}

\maketitle

\section{Introduction}

It is well known that Carnot efficiency is the upper bound for the efficiency of heat engines operating between two reservoirs at different temperatures. However, the heat engines working at Carnot efficiency output zero power. It is essential for the engines to produce sufficient power in practice, thus the Carnot cycle should be speeded up and performed in finite time. Since the seminal achievements were made by Yvon \cite{Yvon55}, Novikov \cite{Novikov}, Chambadal \cite{Chambadal}, Curzon and Ahlborn \cite{Curzon1975}, the problem of efficiency at maximum power (EMP) for heat engines has been attracted much attention \cite{Andresen77,Rubin79,Salamon80,Devos85,Chen1989,ANGULOBROWN91,ChenJC94,Bejan96,ChenL99,vdbrk2005,SanchoPRE06,dcisbj2007,ChenJNET2011,Moreaupre12,TuCPB12,WangHe12,RocoPRE12,Schmiedl2008,Tu2008,Esposito2009a,Esposito2009,Izumida,wangx10,wangx11}.

Different model systems exhibit quite different behaviors at large relative temperature difference between two thermal reservoirs although they show certain universal behavior at small relative temperature difference \cite{Schmiedl2008,Tu2008,Esposito2009a,Esposito2009,wangx11}, which leads to recent investigations on the bounds of
EMP for Carnot-like heat engines \cite{EspositoPRE10,Esposito2010,Velasco10,GaveauPRL10,WangTu2011}. Esposito \emph{et al.} proposed the model of low-dissipation Carnot-like heat engines inspired by their previous work on EMP of quantum-dot Carnot engines \cite{EspositoPRE10} and found that the EMP of low-dissipation Carnot-like heat engines is bounded between $\eta_C/2$ and $\eta_C/(2-\eta_C)$ \cite{Esposito2010}, where $\eta_C$ is the Carnot efficiency. S\'{a}nchez-Salas \emph{et al.} derived EMP to be bounded between $\eta_C/2$ and $\eta_C(1+\eta_C)/2$ by assuming that all coefficients in the Taylor series expansion of EMP with respect to $\eta_C$ are positive \cite{Velasco10}. Gaveau and his coworkers proposed a novel definition of efficiency (the sustainable efficiency) and proved that the sustainable efficiency has the upper bound $1/2$, based on which they also obtained the upper bound ${\eta_C}/(2-\eta_C)$ for the EMP of Carnot-like engines \cite{GaveauPRL10}. The present authors investigated Carnot-like heat engines within the framework of linear irreversible thermodynamics and also found that the EMP of Carnot-like heat engines has the same bounds as those obtained by Esposito and his coworkers \cite{WangTu2011}. Seifert argued that the upper bound $1/2$ for the sustainable efficiency holds only in the linear nonequilibrium regime \cite{Seifert2011}, therefore the upper bound $\eta_C/(2-\eta_C)$ might not exist for EMP of Carnot-like heat engines in the regime far away from equilibrium \cite{Seifert2011p}.

In this Letter, we address the issue of EMP for irreversible Carnot-like heat engines, which is based on the weak endoreversible assumption and the phenomenologically irreversible
thermodynamics. Here the word ``weak" means that the effective temperature of working substance is not presumed to be constant in the finite-time ``isothermal" processes of Carnot-like heat engines, which is different from the conventional endoreversible assumption where the effective temperature of working substance is presumed to be constant in the isothermal processes. The quotation marks on ``isothermal" merely indicate that the working substance is in contact with a thermal reservoir at constant temperature. It is found that the Carnot-like heat engines working at maximum power require the irreversible entropy production in each
finite-time ``isothermal" processes to reach the minimum for given time intervals, which further results in that the effective temperature of working substance in each ``isothermal" process happens to be constant for the engines working at maximum power. Thus the weak endoreversible assumption reduces to the conventional one for the heat engines working at maximum power. Additionally, we classify the Carnot-like heat engines into three types (linear, superlinear, and sublinear) according to the characteristics of constitutive relations between the heat transfer rate and the thermodynamic force.
The EMPs of linear, superlinear, and sublinear irreversible Carnot-like heat engines are found to be bounded between $\eta_C/2$ and $\eta_C/(2-\eta_C)$, 0 and $\eta_C/(2-\eta_C)$, and $\eta_C/2$ and $\eta_C$, respectively. The above classifications and bounds are confirmed by two concrete examples.

\section{Model}
The heat engines that we concerned perform Carnot-like cycle consisting of four steps as follows.

\subsection{``Isothermal" expansion} The
working substance expands in contact with a hot reservoir at
temperature $T_1$ and absorbs heat $Q_1$ from the hot reservoir
during the time interval $0<\tau<t_1$ where $\tau$ is a time variable. The variation of entropy in this process can be expressed as
\begin{equation}
\Delta S_1=\frac{Q_1}{T_1} +\Delta S_1^{ir},
\label{Eq-deltaS1}
\end{equation}
where $\Delta S_1^{ir}\geq 0$ is the irreversible entropy production.

\subsection{Adiabatic expansion} The
working substance decouples from the hot reservoir at time $\tau=t_1$ and then expands without any heat exchange during time interval $t_1<\tau<t_1+t_2$. At time $\tau=t_1+t_2$, the working substance is in contact with a cold
reservoir. According to the convention adopted by many researchers \cite{Curzon1975,Andresen77,Rubin79,Salamon80,Devos85,Chen1989,vdbrk2005,SanchoPRE06,dcisbj2007,Esposito2009,Esposito2010}, the entropy production in this process is regarded as $\Delta S_2=0$.

\subsection{``Isothermal" compression} The working substance is compressed in contact with the cold
reservoir at temperature $T_3$ and releases heat $Q_3$ to the cold reservoir during time interval $t_1+t_2<\tau<t_1+t_2+t_3$. The variation of entropy in this process can be expressed as
\begin{equation}
\Delta S_3 = - \frac{Q_3}{T_3}+\Delta S_3^{ir},
\label{Eq-deltaS3}
\end{equation}
where $\Delta S_3^{ir}\geq 0$ is the irreversible entropy production.

\subsection{Adiabatic compression} The
working substance decouples from the cold reservoir at time $\tau=t_1+t_2+t_3$ and is further compressed without any heat exchange during time interval $t_1+t_2+t_3<\tau<t_1+t_2+t_3 +t_4$. At time $\tau=t_1+t_2+t_3 +t_4$, the working substance is in contact with the hot reservoir again and recovers to its initial state. In this process, both the heat exchange and the entropy production are vanishing, i.e. $Q_4=0$ and $\Delta S_4=0$.

\section{Assumptions}
To continue our analysis, we take two key assumptions as follows.

\subsection{Total time assumption} Following Curzon and Ahlborn \cite{Curzon1975}, we assume the total time for completing the whole cycle is proportional to the time for completing two ``isothermal" processes, i.e., $t_{_{tot}}=t_1+t_2+t_3+t_4=\xi(t_1+t_3)$ with a constant parameter $\xi$.

\subsection{Weak endoreversible assumption} The relaxation time scale for the motion of molecules of working substance is much smaller than the time scale of heat exchange between molecules and the reservoirs in the ``isothermal" processes. Thus the working substance can easily relax to endoequilibrium (i.e., internal equilibrium) in the time scale of the heat exchange. It is in this sense that we can introduce the effective temperatures $T_{1e}$ and $T_{3e}$ of the working substance when it contacts with the hot reservoir or the cold one, respectively. The effective temperatures are not presumed to be constant because they usually depend on the detailed protocols such as the moving speed of the piston. The time-dependent effective temperature is defined in the sense that the
infinitesimal time element ¡°$\mathrm{d}\tau$¡± is still much larger than the relaxation time scale for the
motion of molecules. This assumption is called weak endoreversible assumption because
within the framework of irreversible thermodynamics, this assumption implies that the heat
engines working at maximum power happen to satisfy the original endoreversible relation [eq.~(\ref{convendorev}) in this Letter] proposed in Ref.~\cite{Curzon1975}.

\section{Optimizing the power}
Having undergone a whole cycle, the system recovers to its initial state. Thus the changes of total energy and entropy (both are state functions) are vanishing, from which we can easily derive the variations of entropy $\Delta S_1 =-\Delta S_3 \equiv \Delta S$ in two ``isothermal" processes and the net work output $W = Q_1-Q_3$ in the whole cycle. Having considered eqs.(\ref{Eq-deltaS1}) and (\ref{Eq-deltaS3}), we obtain the power
\begin{equation}
P \propto \frac{Q_1 -Q_3}{t_1+t_3}=\frac{(T_1-T_3)\Delta S-(T_1\Delta S_1^{ir} + T_3 \Delta S_3^{ir})}{t_1+t_3}.
\label{Eq-powerWT}
\end{equation}
Because $\Delta S$ is a state variable only depending on the initial and final states of the ``isothermal" processes while $\Delta S_1^{ir}$ and $\Delta S_3^{ir}$ are process variables depending on the detailed protocols, it is easy to realize that maximizing power implies minimizing $\Delta S_1^{ir}$ and $\Delta S_3^{ir}$ with respect to the protocols for given time intervals $t_1$ and $t_3$.

In the ``isothermal" expansion process, the thermodynamic force may be expressed as $F_1=1/T_{1e}(\tau)-1/T_{1}$ \cite{Prigogine} while the heat transfer rate can be formally expressed as $q_1=q_1(F_1)$. Thus the rate of irreversible entropy production can be written as $\sigma_1=q_1(F_1) F_1$. To minimize
$\Delta S_1^{ir}=\int_0^{t_1}\sigma_1 \mathrm{d}\tau$ with constraint $\int_0^{t_1}q_1(F_1) \mathrm{d}\tau=Q_1$, we need to introduce a Lagrange multiplier $\Lambda_1$ and then minimize the following unconstraint functional
$I=\int_0^{t_1}q_1(F_1) F_1 \mathrm{d}\tau + \Lambda_1[\int_0^{t_1}q_1(F_1) \mathrm{d}\tau-Q_1]$. The corresponding Euler-Lagrange equation
\begin{equation}\label{eq-EL}
(F_{1}+\Lambda_{1})q_{1}^{\prime}+q_{1}(F_{1})  =0
\end{equation}
can be obtained through simple variational calculus,
where $q_{1}^{\prime}$ represents the derivative of $q_{1}$ with respect to $F_{1}$.
Because $\Lambda_{1}$ is independent of the protocol (or time variable $\tau$), the physically acceptable solution is that $F_{1}$
is also independent of time variable $\tau$. We also note that similar conclusion has been drawn by Salamon \emph{et al.} in Ref.~\cite{Salamon80}. Consequently, the effective temperature of working substance happens to be constant in the ``isothermal" expansion process when the heat engine is working at maximum power, that is, the thermodynamic force can be simply written as $F_1=1/T_{1e}-1/T_{1}$ without time variable $\tau$ in this case. Then
we have $Q_1 =\int_0^{t_1}q_1(F_1) \mathrm{d}\tau =q_1 t_1$ and $\Delta S_1^{ir} =\int_0^{t_1}q_1(F_1) F_1 \mathrm{d}\tau= q_1(F_1) F_1 t_1 =Q_1 F_1$ for the heat engine working at maximum power. Substituting these formulas into eq.~(\ref{Eq-deltaS1}), we arrive at $\Delta S_1={Q_1}/{T_{1e}}$.

Similarly, for the ``isothermal" compression process, we can prove that the effective temperature $T_{3e}$ of working substance, the thermodynamic force $F_3=1/T_{3}-1/T_{3e}$, and the heat transfer rate $q_3=q_3(F_3)$ are also independent of time variable $\tau$ for the heat engine working at maximum power. Then we can further obtain $Q_3 =q_3 t_3$ and $\Delta S_3^{ir} =Q_3 F_3$ for the heat engine working at maximum power. Substituting these formulas into eq.~(\ref{Eq-deltaS3}), we arrive at $\Delta S_3=-{Q_3}/{T_{3e}}$. Considering the relation $\Delta S_3=-\Delta S_1$, we arrive at \begin{equation}\label{convendorev}
\frac{Q_1}{T_{1e}}=\frac{Q_3}{T_{3e}}
\end{equation}
for the heat engine working at maximum power. The above equation is no more than the conventional endoreversible assumption proposed by Curzon and Ahlborn in their classic work \cite{Curzon1975}. So far, we have obtained the first main result in this Letter: the weak endoreversible assumption can reduce to the conventional one [i.e., eq.~(\ref{convendorev})] for the heat engine working at maximum power. The weak endoreversible assumption is more general than the conventional one because we need not presume the effective temperature of working substance in each ``isothermal" process to be constant. It is the requirement of maximum power that happens to confine the effective temperature of each ``isothermal" process in certain value such that the conventional endoreversible assumption holds by chance.

For the sake of convenience, we introduce notations $\alpha = 1/T_{1}$, $\alpha_e = 1/T_{1e}$, $\beta=1/T_3$, $\beta_e=1/T_{3e}$, $F_1=\alpha_e-\alpha$, and $F_3=\beta-\beta_e$. Heat engines absorb heat from the hot reservoir and output work, then a certain amount of heat is released into the cold reservoir, which requires $T_{1}>T_{1e}>T_{3e}>T_3$, that is,
$\beta>\beta_e>\alpha_e >\alpha$, $F_1>0$ and $F_3 >0$. With these notations, the Carnot efficiency can be expressed as $\eta_C =1-\alpha/\beta$. By considering eq.~(\ref{convendorev}), the efficiency of Carnot-like heat engines can be derived as
\begin{equation}\label{etaCLHE}
\eta = \frac{Q_1 -Q_3}{Q_1} = 1-\frac{\alpha_e}{\beta_e}.
\end{equation}
Of course, $\eta$ should be bounded between $0$ and $\eta_C$ for heat engines, which can also be derived from the inequality $\beta>\beta_e>\alpha_e >\alpha$ mentioned above. It is of great interest and significance to discuss whether there exists more precise bounds of efficiency when the heat engines working at maximum power.

Noting that $Q_1 =q_1 t_1$, $Q_3 =q_3 t_3$ and eq.~(\ref{etaCLHE}), we can derive the power
\begin{equation}P\propto\frac{Q_{1}-Q_{3}}{t_{1}+t_{3}}=\frac{\eta q_{1}q_{3}}{(1-\eta)q_{1}+q_{3}}.
\label{Eq-powerWT2}
\end{equation}
Now if we take $\beta_e$ and $\eta$ as independent variables, the other variables can be expressed as $F_{3}  =\beta-\beta_{e}$, $q_{3}=q_{3}(F_{3})$, $F_{1}=\alpha_{e}-\alpha=(1-\eta)  \beta_{e}-\alpha$ and $q_{1}=q_{1}(F_{1})$. From $\partial P/\partial \beta_{e} =0$ and $\partial P/\partial \eta =0$, we derive
\begin{equation}\label{key-eq1}
\frac{q_{3}^{\prime}}{q_{3}^{2}}=\frac
{q_{1}^{\prime}}{q_{1}^{2}},
\end{equation}
and
\begin{equation}\label{key-eq2}
\beta_{e}\eta=\frac{q_{1}}{q_{1}^{\prime}}+\frac{q_{1}^2}{q_{3}q_{1}^{\prime}}
\end{equation}
respectively, where $q_{3}^{\prime}$ represents the derivative of $q_{3}$ with respect to $F_{3}$.
Substituting eq.~(\ref{key-eq1}) into (\ref{key-eq2}), we obtain
\begin{equation}\label{key-eq0}
\beta_{e}\eta=\frac{q_{1}}{q_{1}^{\prime}}+\frac{q_{3}}{q_{3}^{\prime}}.
\end{equation}
This is our key equation and the second main result in this Letter. All the following discussions are based on this key equation.

\begin{figure}[!htp]\begin{center}
\includegraphics[width=7.5cm]{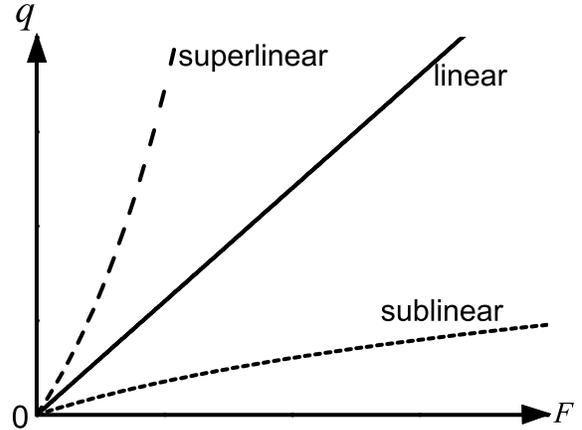}
\caption{\label{fig-3type} Schematic diagram of three irreversible types of constitutive relations.}\end{center}\end{figure}

\section{Classification of constitutive relations}
In the above discussions, the relation between the heat transfer rate and the thermodynamic force is formally expressed as $q=q(F)$ where $q$ represents $q_1$ or $q_3$ while $F$  represents $F_1$ or $F_3$. This relation is called constitutive relation. Intuitively, the constitutive relation can display three kinds of typical characteristics which are schematically depicted in Fig.~\ref{fig-3type}. The first one is called linear type which is represented by the straight line. The second one is called superlinear type which is represented by the convex curve. The third one is called sublinear type which is represented by the concave curve. The behavior of three kinds of constitutive relations can be mathematically expressed as
\begin{equation}\label{eq-3type}
\left\{
                \begin{array}{ll}
                  q/q'=F, & \hbox{linear type;} \\
                  q/q'<F, &\hbox{superlinear type;} \\
                  q/q'>F, & \hbox{sublinear type,}
                \end{array}
              \right.
\end{equation}
where $q^{\prime}$ represents the derivative of $q$ with respect to $F$. For examples, $q=\kappa F+\zeta F^2$ with $\kappa>0$ is of linear, superlinear or sublinear type if $\zeta=0$, $\zeta>0$ or $\zeta<0$, respectively. The constitutive relation of power-law profile $q=\kappa F^{n}$ with $\kappa>0$ is of linear, superlinear or sublinear type if $n=1$, $n>1$ or $0<n<1$, respectively.

\section{Bounds of EMP}
We will discuss the bounds of EMP for three types of Carnot-like heat engines as follows in terms of different kinds of constitutive relations [see eq.~(\ref{eq-3type})] between the heat transfer rate and the thermodynamic force.

\subsection{Linear irreversible engines} The heat transfer rate and the thermodynamic force satisfy the linear relation in two ``isothermal" processes, that is, $q_1/q_1^{\prime}=F_1$ and $q_3/q_3^{\prime}=F_3$ as it is mentioned in eq.~(\ref{eq-3type}).
Substituting them into the above key equation~(\ref{key-eq0}), we find
\begin{equation}\label{key-eqlie1}
\beta_{e}\eta=F_1 +F_3.
\end{equation}
By considering $F_1=\alpha_e-\alpha$, $F_3=\beta-\beta_e$, $\eta_C =1-\alpha/\beta$, eqs.~(\ref{etaCLHE}) and (\ref{key-eqlie1}), we derive
$\beta_{e}={\beta\eta_{C}}/{2\eta}$ and $\alpha_{e} =(1-\eta){\beta\eta_{C}}/{2\eta}$. Because $\beta>\beta_e$ and $\alpha_e>\alpha$, then we derive the lower and upper bounds of EMP to be $\eta_{-}=\eta_C/2$ and $\eta_{+}=\eta_C/(2-\eta_C)$, respectively, for linear irreversible Carnot-like heat engines. This result is consistent with that obtained in our previous work \cite{WangTu2011} based on linear irreversible thermodynamics. A crucial difference is that the bounds are directly derived from eq.~(\ref{key-eqlie1}) in this Letter without calculating the explicit expression of EMP.

\subsection{Superlinear irreversible engines} The heat transfer rate and the thermodynamic force satisfy the superlinear relation in two ``isothermal" processes, that is, $q_j/q_j^{\prime} <F_j$ ($j$=1,3) in terms of eq.~(\ref{eq-3type}). Thus the key equation~(\ref{key-eq0}) is transformed into
\begin{equation}\label{key-eqsuplie1}
\beta_{e}\eta < F_1 +F_3.
\end{equation}
By considering $F_1=\alpha_e-\alpha$, $F_3=\beta-\beta_e$, $\eta_C =1-\alpha/\beta$, eq.~(\ref{etaCLHE}) and inequality~(\ref{key-eqsuplie1}), we derive
$\beta_{e}<{\beta\eta_{C}}/{2\eta}$ and $\alpha_{e} <(1-\eta){\beta\eta_{C}}/{2\eta}$. Given $\alpha_e>\alpha$, then we finally derive the upper bound of EMP to be $\eta_{+}=\eta_C/(2-\eta_C)$ for superlinear irreversible Carnot-like heat engines. The above inequality~(\ref{key-eqsuplie1}) gives no confinement on the lower bound, thus one can take $\eta_{-}=0$ as a conservative estimate.

\subsection{Sublinear irreversible engines} The heat transfer rate and the thermodynamic force satisfy the sublinear relation in two ``isothermal" processes, that is, $q_j/q_j^{\prime} >F_j$ ($j$=1,3) in terms of eq.~(\ref{eq-3type}). Thus the key equation~(\ref{key-eq0}) is transformed into
\begin{equation}\label{key-eqsublie1}
\beta_{e}\eta > F_1 +F_3.
\end{equation}
By considering $F_1=\alpha_e-\alpha$, $F_3=\beta-\beta_e$, $\eta_C =1-\alpha/\beta$, eq.~(\ref{etaCLHE}) and inequality~(\ref{key-eqsublie1}), we derive
$\beta_{e}>{\beta\eta_{C}}/{2\eta}$. Because $\beta>\beta_{e}$, then we finally derive the lower bound of EMP to be $\eta_{-}=\eta_C/2$ for sublinear irreversible Carnot-like heat engines. The above inequality~(\ref{key-eqsublie1}) gives no confinement on the upper bound, thus one can take $\eta_{+}=\eta_C$ as a reasonable estimate.

So far the bounds of EMP ($\eta^\ast$) for three types of irreversible engines can be summarized as
\begin{equation}\label{eq-bnd3type}
\left\{
                \begin{array}{ll}
                  \eta_C/2<\eta^\ast<\eta_C/(2-\eta_C), & \hbox{linear tpye;} \\
                  0<\eta^\ast<\eta_C/(2-\eta_C), &\hbox{superlinear type;} \\
                  \eta_C/2<\eta^\ast<\eta_C, & \hbox{sublinear type;}
                \end{array}
              \right.
\end{equation}
which is the third main result in this Letter.

\section{Examples}
Two concrete examples will help us to understand our classifications and the corresponding bounds of three types of heat engines.

\subsection{Minimally nonlinear relation} For small thermodynamic forces, the minimally nonlinear constitutive relation may be expressed as $q_l=\kappa_lF_l+\zeta_lF_l^2$ ($l=1,3$) where $\kappa_l>0$ and $\zeta_l$ are given parameters. The spirit of this model is similar to the model proposed by Izumida and Okuda \cite{Izumida12}.
It is easy to find that heat engines are of linear, superlinear or sublinear types if $\zeta_l=0$, $\zeta_l>0$ or $\zeta_l<0$, respectively. We will calculate the EMPs for three types of heat engines and then confirm the bounds in eq.~(\ref{eq-bnd3type}).

Substituting the constitutive relation into eqs.~(\ref{key-eq1}) and (\ref{key-eq0}) and then expanding them into Taylor series with respect to $F_1$ and $F_3$, we obtain the following two equations
\begin{eqnarray}\label{Nolinearcase}
    2F_1-\zeta_1/\kappa_1F_1^2 +2F_3-\zeta_3/\kappa_3F_3^2+ \mathcal{O} (F_1^3,F_3^3)=\beta-\alpha, \\
   F_1 =\sqrt{\kappa_3/\kappa_1}F_3+ \mathcal{O} (F_3^3).
\end{eqnarray}
We keep the above equations up to the quadratic terms for small thermodynamic forces $F_1$ and $F_3$, and then derive EMP to be
\begin{equation}\label{NolinearEmp}
    \eta^\ast=\frac{\eta_C}{1+\frac{(1+\sqrt{\kappa_3/\kappa_1}-\eta_C)}{\sqrt{(1+\sqrt{\kappa_3/\kappa_1})^2-\beta\eta_C(\zeta_3/\kappa_3+\zeta_1\kappa_3/\kappa^2_1)}}}.
\end{equation}

If we considering the linear irreversible heat engines, $\zeta_l=0$, the above equation degenerates into $\eta_{CY}=\eta_C/[2-\eta_C/(1+\sqrt{\kappa_3/\kappa_1})]$ obtained by Chen and Yan \cite{Chen1989}. Obviously, the EMP is bounded between $\eta_C/2$ and $\eta_C/(2-\eta_C)$.

Considering the superlinear irreversible heat engines, $\zeta_l>0$, we can easily drive $\eta^\ast<\eta_C/[2-\eta_C/(1+\sqrt{\kappa_3/\kappa_1}~)]<\eta_C/(2-\eta_C)$ from eq.~(\ref{NolinearEmp}). In addition, equation~(\ref{NolinearEmp}) implies that the difference between the lower bound of EMP and $\eta_C/2$ can be expressed as
\begin{equation}\label{superlowbd1}\eta_{-}-\eta_C/2 =- (\beta \zeta_1/8\kappa_1)\eta_C^2 +\mathcal{O}(\eta_C^3),\end{equation}
which implies that the EMP of superlinear irreversible heat engines can be smaller than $\eta_C/2$, and that the stronger nonlinearity results in even smaller lower bound. Thus we take $0$ as the universally lower bound of EMP for superlinear irreversible heat engines.

On the other hand, $\zeta_l<0$ for the sublinear irreversible heat engines. In this case we can derive $\eta^\ast>\eta_C/[2-\eta_C/(1+\sqrt{\kappa_3/\kappa_1}~)]>\eta_C/2$ from eq.~(\ref{NolinearEmp}). Similarly, the difference between the upper bound of EMP and $\eta_C/(2-\eta_C)$ can be expressed as
\begin{equation}\label{subupbd1}\eta_{+}-\eta_C/(2-\eta_C) =- (\beta \zeta_3/8\kappa_3)\eta_C^2 +\mathcal{O}(\eta_C^3).\end{equation} Because $\zeta_3<0$, the above equation implies that the EMP of sublinear irreversible heat engines can be larger than $\eta_C/(2-\eta_C)$, and that the stronger nonlinearity results in even larger upper bound. Thus we take $\eta_C$ as the universally upper bound for sublinear irreversible heat engines.

\subsection{Power-law profile} We mathematically consider simple constitutive relation of power-law profile, $q_l=\kappa_l F_l^{n}$ ($l=1,3$) where $\kappa_l >0$ and $n>0$ are given parameters. It is easy to find that heat engines are of linear, superlinear or sublinear types if $n=1$, $n>1$ or $0<n<1$, respectively. We note that the Dulong-Petit's heat transfer law \cite{AngulBrown93jap} degenerates into the power-law profile with $n\ge 1$ for small thermodynamic forces. Substituting $q_l=\kappa_l F_l^{n}$ into eqs.~(\ref{key-eq1}) and (\ref{key-eq0}), we can explicitly derive the EMP to be
\begin{equation}\label{emp}
    \eta^\ast=\frac{\eta_C}{(n+1)-n \eta_C/[1+(\kappa_3/\kappa_1)^{1/(n+1)}]},
\end{equation}
from which we find that
\begin{equation}\label{empbund}
   \eta_C/(n+1)< \eta^\ast <\eta_C /(n+1-n\eta_C).
\end{equation}

For the linear irreversible engines, $n=1$, eq.~(\ref{empbund}) implies $\eta_C/2<\eta^\ast<\eta_C/(2-\eta_C)$.
For the superlinear irreversible engines, $n>1$, eq.~(\ref{empbund}) implies $0\leq \eta_C/(n+1)< \eta^\ast <\eta_C /(n+1-n\eta_C)\leq\eta_C/(2-\eta_C)$ where the lower bound $0$ can be reached for sufficiently large $n$ while the upper bound can be reached for $n\rightarrow 1^{+}$. For the sublinear irreversible engines, $0<n<1$, eq.~(\ref{empbund}) implies $\eta_C/2\leq\eta_C/(n+1)< \eta^\ast <\eta_C /(n+1-n\eta_C)\leq \eta_C$ where the upper bound $\eta_C$ can be reached for sufficiently small $n$ while the lower bound can be reached for $n\rightarrow 1^{-}$.

\section{Conclusion and discussion}
The issue of EMP for Carnot-like heat engines is investigated based on the weak endoreversible assumption and the phenomenologically irreversible thermodynamics. The heat engines are classified into three irreversible types according to the characteristics of constitutive relations. The bounds of EMP for three types of Carnot-like heat engines are obtained and display certain universality for each types [see eq.~(\ref{eq-bnd3type})]. These results and two examples improve our understanding of the issues of EMP for heat engines and irreversible thermodynamics. There still remains several points as follows which need to be further clarified.

(i) We note that the problem on the bounds of EMP for Carnot-like heat engines comes down to judging the comparative relation between the right-handed side term of our key equation~(\ref{key-eq0}) and the sum of thermodynamic forces. Three types of heat engines mentioned above present definite relations. However, there still exists a mixed irreversible type of heat engines with different kinds of constitutive relations in two ``isothermal" processes (the sublinear constitutive relation in the ``isothermal" expansion process and the suplinear one in the ``isothermal" compression process, or vice versa). The Curzon-Ahlborn engine \cite{Curzon1975} is a typical representative of this type.
Because this type of engines gives indefinite relation between the right-handed side term of eq.~(\ref{key-eq0}) and the sum of thermodynamic forces, we cannot simply obtain the universal bounds of EMP based on eq.~(\ref{key-eq0}).
However, we can still obtain the explicit expression of EMP by considering eqs.~(\ref{key-eq1}) and (\ref{key-eq0}) simultaneously. After we transform the heat transfer law $q_1= \alpha (T_1- T_{1e})$ and $q_3= \beta (T_{3e}- T_{3})$ used by Curzon and Ahlborn into $q_1=\alpha T_1^2 F_1/(1+T_1F_1)$ and $q_3=\beta T_3^2 F_3/(1-T_3F_3)$ in terms of the thermodynamic forces $F_1$ and $F_3$, we can easily obtain $\eta^\ast=1-\sqrt{T_3/T_1}$ from eqs.~(\ref{key-eq1}) and (\ref{key-eq0}).

(ii) In Ref.~\cite{Izumida}, the effective temperature of gas molecules in a cylinder is investigated when the piston moves at a constant velocity. It is found that the effective temperature exhibits very strange behavior in a small region where the working substance is switched from ``isothermal" process to adiabatic one. The size of this region should depend on the interactions among gas molecules and those between gas molecules and reservoirs. If the proper interaction parameters are selected such that the relaxation time scale for the motion of molecules of working substance is much smaller than the time scale of heat exchange between molecules and the reservoirs, we expect that the weak endoreversible assumption can hold and the size of the strange region can be neglected. In addition, the engines working at maximum power might take the protocol that corresponds to the nonuniform speed of piston. Under this protocol, the strange region might be reduced, which will be investigated in our future work.

(iii) We have arrived at the result that the heat transfer rate is time-independent if the effective temperature of working substance is constant, which is consistent with the macroscopic heat transfer law adopted by many researchers \cite{Curzon1975,Andresen77,Rubin79,Salamon80,Devos85,Chen1989,ANGULOBROWN91,ChenJC94,Bejan96,ChenL99}. However, Schmiedl and Seifert \cite{Schmiedl2008} observed that the heat transfer rate in a microscopic model is time-dependent although the effective temperature of the Brownian particle is constant. As we know, this paradox between the macroscopic and microscopic models is still an open question. Our key results in this Letter might only be applicable for the macroscopic heat engines. We will address this paradox by extending the constitutive relation $q=q(F(\tau))$ which only depends implicitly on the time variable to the form $q=q(F(\tau),\tau)$ that also depends explicitly on the time variable in the future work.

(vi) The entropy productions for both adiabatic processes are presumed to be zero in almost all extant models. In microscopic models, the vanishing entropy productions can indeed be realized by instantaneous adiabatic transitions \cite{Schmiedl2008,EspositoPRE10}. However, we still have limited information in whether or how we can select a proper protocol to guarantee the vanishing entropy production in finite-time adiabatic processes for macroscopic models, which might constitute one of major challenges in the finite-time thermodynamics.

\acknowledgments
The authors are grateful to the financial support from Nature Science Foundation of China (Grant NO. 11075015). ZCT is also grateful to Udo Seifert for his instructive discussions.

\end{document}